\documentclass[superscriptaddress,amsmath, nofootinbib]{revtex4}
\usepackage{amsfonts}
\usepackage{graphicx}
\usepackage[brazil]{babel}
\usepackage[latin1]{inputenc}
\usepackage{amsmath}
\usepackage{amssymb}

\begin{document}


\title{Covariant version of the Pauli Hamiltonian, spin-induced noncommutativity, Thomas precession and precession of spin.}

\author{Alexei A. Deriglazov }
\email{alexei.deriglazov@ufjf.edu.br} \affiliation{Depto. de Matem\'atica, ICE, Universidade Federal de Juiz de Fora,
MG, Brazil} \affiliation{Department of Physics, Tomsk State University, Lenin Prospekt 36, 634050, Tomsk, Russia}

\author{Danilo Machado Tereza }
\email{danilomachadot@gmail.com} \affiliation{Depto. de Matem\'atica, ICE, Universidade Federal de Juiz de Fora, MG,
Brazil}

\date{\today}

\begin{abstract}
We show that there is a manifestly covariant version of the Pauli Hamiltonian with equations of motion quadratic on spin and field strength. Relativistic covariance inevitably leads to noncommutative positions: classical brackets of the position variables are proportional to the spin. It is the spin-induced noncommutativity that is responsible for transforming the covariant Hamiltonian into the Pauli Hamiltonian, without any appeal to the Thomas precession formula. The Pauli theory can be thought to be $1/c^2$\,-approximation of the covariant theory written in special variables. These observations clarify the long standing question on the discrepancy between the covariant and Pauli Hamiltonians. We also discuss the transformational properties of spin axis in the passage from laboratory to comoving and instantaneous frames, and reveal the role of Thomas spin-vector in the covariant scheme.
\end{abstract}

\maketitle 

\section{Introduction. Relativistic spinning electron and the problem of covariant formalism.}\label{pau}

Classical models of relativistic spin represent a working tool used to describe the behavior of elementary particles and rotating bodies in electromagnetic and gravitational fields.
One obscure point of this approach, which has been raised for discussion already in the pioneer works \cite{Gousmith_1926, Thomas1927, Frenkel, Frenkel2} and remains under debates up to date,
 is the so-called problem of covariant formalism.
Clarification  of this issue could be of interest in various areas, including muon and electron $g-2$\,-experiments \cite{Field:1979, miller2007muon, Struck_2019, Dayi_2015}, influence of spin on the trajectory of a rotating body in general relativity \cite{Khripl_2000, Oancea_2019, Gerakopoulos_2019, Bosso_2019, Nica_2019, Dehghani_2018}, and black hole physics near horizon \cite{Khripl_2000, Nagar_2018_25, Kaye_2019, Bosso_2018, Peng_2019}.
In the textbooks and reviews, it has become almost a tradition to discuss the problem without formulating it in an exact form\footnote{See p. 541 and 558 in \cite{Jackson}, p. 87 in \cite{Sak}, p. 6 in \cite{Weinb_QFT}, p. 563 in \cite{Field:1979}.}. Let us try to break this tradition.

Historically, the notion of a classical spinning electron \cite{Gousmith_1926, Thomas1927, Frenkel, Frenkel2} has been developed in attempts to explain the energy levels of
atomic spectra. Following the ideas of Uhlenbeck and Goudsmit \cite{Gousmith_1926}, Thomas accepted that  spinning particle can be described using its position vector ${\bf x}(t)$, and the vector of
spin-axis ${\bf S}_T(t)$ attached to the particle. The position vector obeys the Lorentz-force equation
\begin{eqnarray}\label{pau.1}
m\ddot{\bf x}=e{\bf E}+\frac{e}{c}[\dot{\bf x}, {\bf B}],
\end{eqnarray}
while the rate of variation of spin was initially assumed to be
\begin{eqnarray}\label{pau.2}
\frac{d{\bf S}_T}{dt}=-\frac{e}{mc}\left\{[{\bf B}, {\bf S}_T]-\frac{1}{2mc}[[{\bf p}, {\bf E}],{\bf S}_T]\right\}.
\end{eqnarray}
We use the notation $[{\bf A}, {\bf B}]$ and $({\bf A}, {\bf B})$ for the vector and scalar products of three-dimensional vectors. 
The quantity ${\bf S}_T$ will be called the Thomas spin-vector.
Here ${\bf E}=\alpha{\bf x}/|{\bf x}|^3$ is Coulomb electric field and ${\bf B}$ is a constant magnetic field.
Thomas showed \cite{Thomas1927} that these equations, together with the Bohr quantization
rule of angular momentum, give a satisfactory description of atomic energy levels. A more systematic calculation of the
energies was achieved in quantum mechanics constructed on the base of these equations. Pauli noticed \cite{Pauli_1927} that Eqs.
(\ref{pau.1}) and (\ref{pau.2}) follow from the Hamiltonian
\begin{eqnarray}\label{pau.3}
H=\frac{1}{2m}({\bf p}-\frac{e}{c}{\bf A})^2+eA^0-\frac{e}{mc}\left[({\bf S}_T, {\bf B})+\frac{1}{2mc}({\bf S}_T, [{\bf E},
{\bf p}])\right],
\end{eqnarray}
with use of canonical brackets (we show the nonvanishing brackets)
\begin{eqnarray}\label{pau.4}
\{x^i, p^j\}=\delta^i{}_j, \qquad \{S_T^i, S_T^j\}=\epsilon^{ijk}S_T^k.
\end{eqnarray}
So, in the Pauli formalism it is assumed that (\ref{pau.2}) represents the equation of motion of spin {\it in the laboratory
system}. Having at hand the Hamiltonian formulation, Pauli constructed quantum mechanics of the spinning electron by
replacing the classical variables $z^A\equiv( x^i, p^k, S^j_T)$ by operators that, according to Dirac quantization rule, must
obey the commutators resembling the classical brackets
\begin{eqnarray}\label{pau.5}
[\hat z^A, \hat z^B]=i\left.\hbar\{z^A, z^B\}\right|_{z\rightarrow \hat z} .
\end{eqnarray}
The operators are
$\hat p^i=-i\hbar\partial_i,  \hat x^i=x^i, \hat S^i=\frac{\hbar}{2}\sigma^i$,
where $\sigma^i$ are $2\times 2$ matrices of Pauli. The operators act on the space of two-component wave functions
$\Psi_a(t, {\bf x})$, $a=1, 2$.  Replacing classical variables in Eq. (\ref{pau.3}) by the operators, he obtained
quantum Hamiltonian and showed that the resulting quantum mechanics reproduces the atomic energy levels \cite{Pauli_1927}. The Hamiltonian (\ref{pau.3}) can also be obtained from the Dirac equation \cite{Dirac_1927, foldy:1978}.

Difficulties arose when trying to develop the manifestly covariant relativistic generalization of the classical theory (\ref{pau.1})-(\ref{pau.3}). Numerous
attempts (see the pioneer works \cite{Thomas1927, Frenkel, Frenkel2, Kramers_1934, BMT_1959} and the reviews \cite{corben:1968, Khripl_2000})  lead to equations of spin and to Hamiltonians that in $1/c^2$\,-approximation differ from those of Pauli theory. For instance,
assuming that three-dimensional spin is a spatial part of a manifestly covariant four-dimensional spin-tensor of Frenkel \cite{Frenkel, Frenkel2},
the covariant theory implies the following expressions (for the details, see below):
\begin{eqnarray}\label{pau.7}
\frac{d{\bf S}}{dt}=-\frac{e}{mc}\left\{[{\bf B}, {\bf S}]+\frac{1}{mc}[{\bf E},[{\bf p}, {\bf S}]]\right\},
\end{eqnarray}
\begin{eqnarray}\label{pau.8}
H_{ph}=\frac{1}{2m}({\bf p}-\frac{e}{c}{\bf A})^2+eA^0-\frac{e}{mc}\left[({\bf S}, {\bf B})+\frac{1}{mc}({\bf S}, [{\bf
E}, {\bf p}])\right].
\end{eqnarray}
We call the quantity ${\bf S}$ the Frenkel spin-vector.
The Hamiltonians (\ref{pau.3}) and (\ref{pau.8}) differ by the famous $1/2$\,-factor in front of the last term, whereas
the last terms in the equations of spin differ in a structure. The question, why a covariant formalism does not lead
directly to the expected result, was raised already in 1926 \cite{Thomas1927, Frenkel, Frenkel2} and remain under discussion to date; see the review \cite{Khripl_2000}.

In attempts to explain the discrepancy, Thomas compared the variation rates of spin axis in comoving and instantaneous frames. The resulting relation is the famous formula of Thomas
precession. It should be noted that the Thomas formula itself is the object of numerous debates. A detailed analysis of controversial works on the subject can be found in \cite{stepanov2012, Malykin_2007, Ritus_2007}. In particular, it is widely believed \cite{Malykin_2007, Field:1979, Jackson, Sak, Weinb_QFT}, that Thomas precession is relevant to the problem of covariant formalism.
In this regard, we note that the Thomas formula relates quantities of different coordinate systems, while the equations
(\ref{pau.1})-(\ref{pau.3}), (\ref{pau.7}), (\ref{pau.8}) are taken in the same (laboratory) system.

The aim of this work is to clarify these issues. We present a manifestly-covariant formulation of a spinning particle, that in $1/c^2$\,-approximation implies the quantum mechanics of Pauli without any appeal to the Thomas precession formula.  Then we make a detailed comparison of Thomas and Frenkel spin-vectors.

The work is organized as follows. In Sect. \ref{mch} we start from the covariant formalism of a spinless particle, and, assuming that spin in a relativistic theory can be described by the Frenkel spin-tensor, we write the expected expressions for manifestly covariant and physical Hamiltonians of a spinning particle. In Sect. \ref{mcf} we fix the brackets that guarantee the consistency of Hamiltonian equations with the supplementary condition on the Frenkel spin-tensor. In particular, we show that the position variables turn out to be noncommuting in the covariant scheme. In Sect. \ref{pfd} we show, how this spin-induced noncommutativity of positions solve the problem of covariant formalism. The calculations in these sections are valid in $1/c^2$\,-approximation.  In Sect. \ref{mcv} we construct the fully covariant brackets and show that there exists the manifestly covariant version of Pauli theory with the equations of motion no more than quadratic on spin and field strength.
In Sect. \ref{thp} return back to the discussion of covariant theory in $1/c^2$\,-approximation, and reveal the meaning of Thomas spin in the covariant scheme.

\section{Manifestly covariant and physical-time Hamiltonians.}\label{mch}

Here we review the known formalism of a spinless particle \cite{deriglazov2010classical}  and then discuss the most natural way to include spin into the covariant scheme. To work with manifestly-covariant expressions, we describe trajectory of the particle in a parametric form:
${\bf x}(t) \rightarrow x^\mu(\tau)=( ~ct(\tau),  {\bf x}(\tau)\equiv{\bf x}(t(\tau)~)$.
As the parameter $\tau$ we take the proper time
\begin{eqnarray}\label{ptp.1}
\tau(t)=\int_0^t dt\sqrt{1-{\bf v}^2/c^2}, \qquad \mbox{then} \quad \left.\frac{dx^0(\tau)}{d\tau}\right|_{\tau(t)}=c\gamma,
\quad \left.\frac{d{\bf x}(\tau)}{d\tau}\right|_{\tau(t)}=\gamma\frac{d{\bf x}(t)}{dt},
\end{eqnarray}
where ${\bf v}(t)=\frac{d{\bf x}}{dt}$,  $\gamma=(1-{\bf v}^2/c^2)^{-\frac12}$, and by construction of the proper time, tangent vector $u^\mu\equiv dx^\mu(\tau)/d\tau$ to the curve $x^\mu(\tau)$ has fixed length at any instant, $\left(u^\mu\right)^2=-c^2$. In accordance with Eq. (\ref{ptp.1}), the four and three-dimensional vectors of velocity are related as follows:
$(u^0, {\bf u})|_{\tau(t)}=( ~c \gamma , ~ \gamma{\bf v} ~)$.
This is a covariant expression, so the last  equation from (\ref{ptp.1}) can be used to restore ${\bf v}(t)$ from ${\bf u}(\tau)$ in any inertial frame.

In the Hamiltonian formulation, to each variable $x^\mu(\tau)$ of configuration space we associate the function $p_\mu(\tau)$ called conjugated momentum. The manifestly covariant Hamiltonian of a spinless particle in electromagnetic field with four-potential $A^\mu$ is
\begin{eqnarray}\label{mch.1}
H=\frac{1}{2m}\left[(p^\mu-\frac{e}{c}A^\mu)^2+(mc)^2\right].
\end{eqnarray}
It is accompanied with the covariant Poisson brackets $\{x^\mu, p_\nu\}=\delta^\mu{}_\nu$,  $\{x^\mu, x^\nu\}=\{p_\mu, p_\nu\}=0$. For
the canonical momentum ${\cal P}^\mu\equiv p^\mu-\frac{e}{c}A^\mu$, the brackets imply $\{{\cal P}^\mu, {\cal
P}^\nu\}=\frac{e}{c}F^{\mu\nu}$. From the requirement of gauge-invariance of the Hamiltonian it follows, that the conjugated momentum $p_\mu$ is not invariant under the $U(1)$\,-gauge transformations
\begin{eqnarray}\label{mch.1.1}
A_\mu=A'_\mu+\partial_\mu\lambda, \quad \mbox{implies}, \quad  p_\mu=p'_\mu+\frac{e}{c}\partial_\mu\lambda.
\end{eqnarray}
In contrast,  ${\cal P}_\mu$ is an invariant object, and we expect that Hamilltonian equations can be presented in terms of this quantity. Computing
$\dot x^\mu=\{x^\mu, H\}$,  $\dot p^\mu=\{p^\mu, H\}$, the equations can be
written in the form
\begin{eqnarray}\label{mch.2}
m\frac{dx^\mu}{d\tau}={\cal P}^\mu, \qquad \frac{d{\cal P}^\mu}{d\tau}=\frac{e}{mc}F^{\mu\nu}{\cal P}_\nu.
\end{eqnarray}
Computing square of the first equation from (\ref{mch.2}), we obtain the mass-shell relation ${\cal P}^2+(mc)^2=0$, or
\begin{eqnarray}\label{mch.3}
cp^0=c\sqrt{(mc)^2+{\boldsymbol{\cal P}}^2}+eA^0.
\end{eqnarray}
Since $x^\mu (\tau)$ and $p_\mu (\tau)$ represent the physical dynamical variables ${\bf x}(t)$ and ${\bf p}(t)$ in a
parametric form, we can write $\frac{d{\bf x}}{dt}=c\frac{d{\bf x}/d\tau}{dx^0 /d\tau}$, $\frac{d{\boldsymbol{\cal
P}}}{dt}=c\frac{d{\boldsymbol{\cal P}}/d\tau}{dx^0 /d\tau}$, and with use of (\ref{mch.2}) and (\ref{mch.3}) we obtain
$\frac{d{\bf x}}{dt}=c{\boldsymbol{\cal P}}/\sqrt{(mc)^2+{\boldsymbol{\cal P}}^2}$, $\frac{d{\boldsymbol{\cal
P}}}{dt}=e{\bf E}+e[{\boldsymbol{\cal P}}, {\bf B}]/\sqrt{(mc)^2+{\boldsymbol{\cal P}}^2}$. Expanding them in series
over $1/c$ and keeping only the terms of order $1/c$, we obtain the well-known equations\footnote{For
electromagnetic field we use the notation:
$F_{\mu\nu}=\partial_\mu A_\nu-\partial_\nu A_\mu=(F_{0i}=-E_i, ~ F_{ij}= \epsilon_{ijk}B_k)$,
$E_i=-\frac{1}{c}\partial_tA_i+\partial_i A_0$, $B_i=\frac12\epsilon_{ijk}F_{jk}=\epsilon_{ijk}\partial_j A_k$.}
\begin{eqnarray}\label{mch.4}
m\frac{d{\bf x}}{dt}={\boldsymbol{\cal P}}, \qquad \frac{d{\boldsymbol{\cal P}}}{dt}=e{\bf E}+\frac{e}{mc}[{\bf p},
{\bf B}].
\end{eqnarray}
The first equation shows that canonical momentum $\boldsymbol{\cal P}$ represents velocity of the particle in Hamiltonian formulation.
Excluding the conjugated momentum ${\bf p}$ from these equations, we arrive at the Lorentz-force equation (\ref{pau.1}).

Equations (\ref{mch.4}) can also be obtained from the Hamiltonian (\ref{mch.1}), presented in terms of the physical
variables ${\bf x}(t)$ and ${\bf p}(t)$. While in general reparametrization-invariant theory with Dirac constraints it requires some caution \cite{Git_2017, Gueor_2019, Hori_2019, Ottinger_2018, Pavsic_2019, Breban_2018, Darvas_2018}, the final result is very simple \cite{DPM2016}: the physical Hamiltonian coincides with the right hand
side of Eq. (\ref{mch.3}).  Its expansion in series up to order $1/c^2$ gives the physical Hamiltonian
\begin{eqnarray}\label{mch.5}
H_{ph}=cp^0\approx mc^2+\frac{1}{2m}({\bf p}-\frac{e}{c}{\bf A})^2+eA^0.
\end{eqnarray}
As it should be, this coincides with spinless part of Pauli Hamiltonian (\ref{pau.3}).

Concerning the spin, we introduce the vector function ${\bf S}(t)$ and take it in the proper-time parametrization,
defining ${\bf S}(\tau)\equiv{\bf S}(t(\tau))$. Following Frenkel \cite{Frenkel}, we identify the components $S^i(\tau)$ of three-dimensional spin with
spatial part of four-dimensional antisymmetric spin-tensor $S^{\mu\nu}=-S^{\nu\mu}$
\begin{eqnarray}\label{mcf.2}
S^{i}=\frac{1}{4}\epsilon^{ijk}S^{jk},  \quad \mbox{then} \quad S^{ij}=2\epsilon^{ijk}S^k.
\end{eqnarray}
We assume that the Frenkel spin $S^i$ can be identified with the spin axis in laboratory frame.
We assume that at each instant of motion, $S^{\mu\nu}$ obeys the covariant condition\footnote{We could equally use
$S^{\mu\nu}p_\nu=0$, with the conjugated momentum $p_\nu$ instead of ${\cal P}_\nu$. The difference between them is of order $1/c$, and does not contribute into subsequent expressions in $1/c^2$\,-approximation.}
\begin{eqnarray}\label{mcf.3}
S^{\mu\nu}{\cal P}_\nu=0, \quad \mbox{then} \quad  S^{0i}=\frac{1}{{\cal P}_0}S^{ij}{\cal P}_j\approx -\frac{2}{mc}[{\bf p}, {\bf S}]^i+
O(1/c^2).
\end{eqnarray}
As a consequence, the number of independent components of spin in relativistic and Pauli theories is the same. In the
rest frame of the particle, where ${\boldsymbol{\cal P}}=0$, the extra-components just vanish, $S^{0i}=0$. Our basic variables $z^A\equiv(x^\mu, p_\nu,
S^{\mu\nu})$ transform linearly under the Lorentz transformations: $x^\mu=\Lambda^\mu{}_\nu x^\nu$,
$p^\mu=\Lambda^\mu{}_\nu p^\nu$, $S^{\mu\nu}=\Lambda^\mu{}_\alpha \Lambda^\nu{}_\beta S^{\alpha\beta}$.

In trying to include spin into the covariant Hamiltonian, we note that the only scalar function containing the desired
spin-field interaction is
$-\frac{e}{2c}F_{\mu\nu}S^{\mu\nu}=-\frac{2e}{c}\left[({\bf S}, {\bf B})+\frac{1}{mc} ({\bf S}, [{\bf E}, {\boldsymbol{\cal P}}])\right]$.
Adding this term to Eq. (\ref{mch.1}), we obtain
\begin{eqnarray}\label{mch.7}
H=\frac{1}{2m}\left[(p^\mu-\frac{e}{c}A^\mu)^2-\frac{e\mu}{2c}F_{\mu\nu}S^{\mu\nu}+(mc)^2\right].
\end{eqnarray}
We added interaction of spin with electromagnetic field through the magnetic moment $\mu$ that corresponds to gyromagnetic ratio $g=2\mu$. In Sections \ref{mcf}, \ref{pfd} and \ref{thp} we will put the classical value of $\mu$: $\mu=1$.
Comparing (\ref{mch.7}) with Eqs. (\ref{mch.1}) and (\ref{mch.5}), the expected expression for physical Hamiltonian in $1/c^2$\, -approximation is
\begin{eqnarray}\label{mch.8}
H_{ph}=
mc^2+\frac{1}{2m}({\bf p}-\frac{e}{c}{\bf A})^2+eA^0-\frac{e}{mc}\left[({\bf S}, {\bf B})+\frac{1}{mc}({\bf S}, [{\bf
E}, {\bf p}])\right].
\end{eqnarray}
We confirm the validity of this expression in Sect. \ref{mcv}.

\section{Non canonical brackets.}\label{mcf}

To obtain equations of motion, the Hamiltonian (\ref{mch.7}) should be accompanied with some brackets. The spin supplementary condition (\ref{mcf.3}) should be consistent with the resulting equations of motion.
This leads us to the
observation that will be crucial for our explanation of $1/2$\,-factor: manifestly-covariant formalism
inevitably leads to relativistic corrections of order $1/c^2$ to the canonical brackets (\ref{pau.4}).

The condition (\ref{mcf.3}) implies $\frac{d}{d\tau}(S^{\mu\nu}{\cal P}_\nu)=0$. In the Hamiltonian formalism, variation rate
of a phase-space function is equal to the bracket of this function with Hamiltonian, so we can write
$\frac{d}{d\tau}(S^{\mu\nu}{\cal P}_\nu)=\{S^{\mu\nu}{\cal P}_\nu, H\}=\{S^{\mu\nu}{\cal P}_\nu, z^A\}\frac{\partial H}{\partial z^A}=0$.
The latter equality certainly holds if\footnote{When this equality is satisfied, we can make the substitution (\ref{mcf.3})
before computing the brackets: $\{K(z), N(z)\}|_{S^{0i}=\frac{1}{{\cal P}_0}S^{ij}{\cal P}_j}=\{K(z)|, N(z)|\}$.}
\begin{eqnarray}\label{mcf.4}
\{z^A, S^{\mu\nu}{\cal P}_\nu\}=0.
\end{eqnarray}
It is this equation that requires a modification of canonical brackets. For $z^A=x^\alpha$ Eq. (\ref{mcf.4})  implies $\{x^\alpha,
S^{\mu\nu}{\cal P}_\nu\}=S^{\mu\nu}\{x^\alpha, {\cal P}_\nu\}+\{x^\alpha, S^{\mu\nu}\}{\cal P}_\nu=0$. This equality holds, if we take
\begin{eqnarray}\label{mcf.4.1}
\{x^\alpha, S^{\mu\nu}\}=\frac{{\cal P}^\mu S^{\nu\beta}-{\cal P}^\nu S^{\mu\beta}}{{\cal P}^2}\{x^\alpha, {\cal P}_\beta\}.
\end{eqnarray}
In turn, if the bracket $\{x^\alpha, {\cal P}_\beta\}=\{x^\alpha, p_\beta\}-\frac{e}{c}\{x^\alpha, x^\gamma\}\partial_\gamma A_\beta$ remains unmodified at the order $1/c^2$, that is $\{x^\alpha, {\cal P}_\beta\}=\delta^\alpha{}_\beta+O(1/c^3)$, the equation (\ref{mcf.4.1}) implies the following modification of canonical bracket $\{x^i, S^j\}$:
\begin{eqnarray}\label{mcf.5}
\{x^i, S^j\}=\frac{p^i S^j-\delta^{ij}({\bf p}, {\bf S})}{(mc)^2},
\end{eqnarray}
at $1/c^2$\,-order.
For the spin-tensor we impose
\begin{eqnarray}\label{mcf.6}
\{S^{\alpha\beta}, S^{\mu\nu}\}=2N^{(\alpha\mu}S^{\beta\nu)}\equiv
2(N^{\alpha\mu}S^{\beta\nu}-N^{\alpha\nu}S^{\beta\mu}-N^{\beta\mu}S^{\alpha\nu}+N^{\beta\nu}S^{\alpha\mu}),
\end{eqnarray}
where $N$ is the projector on the plane orthogonal to ${\cal P}^\mu$:
$N^{\mu\nu}=\eta^{\mu\nu}-\frac{{\cal P}^\mu {\cal P}^\nu}{{\cal P}^2}$, then $N^{\mu\nu}{{\cal P}_\nu}=0$.
The bracket ensures the validity of equation (\ref{mcf.4}) for $z^A=S^{\alpha\beta}$. For the spatial components, Eq. (\ref{mcf.6}) gives
\begin{eqnarray}\label{mcf.8}
\{S^i, S^j\}=\epsilon^{ijk}\left[S^k+\frac{p^k({\bf p}, {\bf S})}{(mc)^2}\right],
\end{eqnarray}
instead of canonical bracket (\ref{pau.4}). The Jacobi identity $\{x^i, \{x^j, S^k\}\}+\{x^j, \{S^k, x^i\}\}+\{S^k, \{x^i, x^j\}\}=0+O(1/c^4)$ with use of (\ref{mcf.5}) requires the following modification of position-position bracket:
$\{x^i, x^j\}=\frac{1}{(mc)^2}\epsilon^{ijk}S^k$.
The examination of Eq. (\ref{mcf.4}) for $z^A=p_\mu$ do not implies $1/c^2$\,-corrections to the canonical brackets of $p_i$ with $x^j$ and $S^k$. So, in  $1/c^2$\,-approximation, the expected nonvanishing brackets of covariant formalism are 
\begin{eqnarray}
\{x^i, x^j\}=\frac{1}{(mc)^2}\epsilon^{ijk}S^k, \label{pfd.5.1} \quad \qquad\\
\{x^i, p^j\}=\delta^i{}_j, \label{pfd.5.2} \qquad \qquad \qquad \quad \\
\{x^i, S^j\}=\frac{p^i S^j-\delta^{ij}({\bf p}, {\bf S})}{(mc)^2}, \qquad \label{pfd.5.3} \\
\{S^i, S^j\}=\epsilon^{ijk}\left[S^k+\frac{p^k({\bf p}, {\bf S})}{(mc)^2}\right]. \label{pfd.5.4}
\end{eqnarray}
We point out that they coincide with $1/c^2$\,-approximation of Dirac brackets in the Ghosh model of anyon \cite{Ghosh_1994}, as well as with $1/c^2$\,-approximation of Dirac brackets arising in the vector model of spin \cite{DPM2016, AAD_Rec}.
Together with the Hamiltonian (\ref{mch.8}), the brackets  (\ref{pfd.5.1}) -(\ref{pfd.5.4})  imply the Frenkel equations (\ref{pau.7}) and (\ref{mch.4}).

\section{The covariant and Pauli formulations determine the same classical and quantum theory.}\label{pfd}
Both $H_{ph}$ written in Eq. (\ref{mch.8}) and the brackets (\ref{pfd.5.1})-(\ref{pfd.5.4}) of covariant theory differ from those of Pauli theory.
Nevertheless, they lead to the same quantum mechanics. Indeed, we can realize our variables by hermitian
operators
\begin{eqnarray}\label{pfd.1}
\hat p^i=-i\hbar\partial_i, \qquad \hat x^i=x^i-\frac{\hbar}{4(mc)^2}\epsilon^{ijk}\hat p^j\sigma^k, \qquad
S^i=\frac{\hbar}{2}\left[\sigma^i-\frac{\hat p^i(\hat{\bf p}, {\boldsymbol\sigma})}{2(mc)^2}+\frac{\hat{\bf
p}^2\sigma^i}{2(mc)^2}\right]\equiv\frac{\hbar}{2}\left[\sigma^i+\frac{1}{2(mc)^2}[\hat{\bf p},[{\boldsymbol\sigma},
\hat{\bf p}]^i\right].
\end{eqnarray}
Their commutators are in correspondence with the classical brackets (\ref{pfd.5.1}) -(\ref{pfd.5.4}), as it should be in accordance with Eq. (\ref{pau.5}).  
We substitute the operators into the Hamiltonian (\ref{mch.8}), and expand the resulting expression in series over
$1/c$ up to $1/c^2$\, -order. In this approximation we have $-\frac{e}{c}{\bf A}(\hat x^i)=-\frac{e}{c}{\bf A}(x^i)+O(1/c^3)$,
while\footnote{Assuming the symmetric ordering of operators, we obtain the hermitian operator
$eA^0(x^i)+\frac{e}{2(mc)^2}(\hat{\bf S},[{\bf E}, \hat{\bf p}])-\frac{i\hbar e}{4(mc)^2}(\hat{\bf S}, \mbox{\bf rot} ~
{\bf E})$. For the central field $\mbox{\bf rot} ~ {\bf E}=0$, and the last term vanishes.}
\begin{eqnarray}\label{pfd.3}
eA^0(\hat x^i)=eA^0(x^i)+\frac{e}{2(mc)^2}(\hat{\bf S},[{\bf E}, \hat{\bf p}]).
\end{eqnarray}
The last term in this expression has the same structure as fourth term in (\ref{mch.8}), so their sum acquires the desired $1/2$\,
-factor. In the result, quantum Hamiltonian of covariant formulation coincides with the Pauli
Hamiltonian
\begin{eqnarray}\label{pfd.4}
\hat H_{ph}= mc^2+\frac{1}{2m}(\hat{\bf p}-\frac{e}{c}{\bf A}(x^i))^2+eA^0( x^i)-\frac{e}{mc}\left[(\hat{\bf S}, {\bf
B})+\frac{1}{2mc}(\hat{\bf S}, [{\bf E}, \hat{\bf p}])\right].
\end{eqnarray}
This solves the problem of covariant formalism.

We also can ask on the relation between covariant and Pauli formulations considered as the classical
theories. Starting from the covariant formulation (\ref{mch.8}), (\ref{pfd.5.1})-(\ref{pfd.5.4}), we look for the
phase-space variables that obey the canonical brackets. They are
\begin{eqnarray}\label{pfd.8}
p_c^i=p^i, \qquad x_c^i=x^i+\frac{1}{2(mc)^2}\epsilon^{ijk}p^jS^k,
\end{eqnarray}
\begin{eqnarray}\label{pfd.9}
S_T^i=S^i+\frac{p^i({\bf p}, {\boldsymbol S})}{2(mc)^2}-\frac{{\bf p}^2S^i}{2(mc)^2}\equiv S^i+\frac{1}{2(mc)^2}[{\bf
p}, [{\bf p}, {\bf S}]]^i.
\end{eqnarray}
The Hamiltonian (\ref{mch.8}) in terms of these variables turns out into the Hamiltonian of Pauli. So at the classical level the
covariant and Pauli formulations are related by (noncanonical) transformation (\ref{pfd.8}), (\ref{pfd.9}) of the
phase-space, and hence describe the same theory\footnote{Note that ${\bf S}_T$ does not coincide with spatial part of Pauli-Lubanski four-vector $S_{PL}^\mu=\frac{1}{4\sqrt{-{\cal P}^2}}\epsilon^{\mu\nu\alpha\beta}{\cal P}_\nu S_{\alpha\beta}$. Its spatial
components $S^i_{PL}=S^i+\frac{1}{(mc)^2}p^i({\bf p}, {\bf S})-\frac{1}{2(mc)^2}{\bf p}^2S^i$ are different from (\ref{pfd.9}).}. In particular, the vector ${\bf S}_T$ defined by (\ref{pfd.9}) should obey the Thomas equation (\ref{pau.2}). It is instructive to show this by direct computation. Using $\frac{d{\bf p}}{dt}=e{\bf E}+O(1/c)$, the Frenkel equation (\ref{pau.7}) can be rewritten as follows: 
\begin{eqnarray}\label{pfd.9.1}
\frac{d}{d\tau}\left[S^i+\frac{1}{2(mc)^2}[{\bf
p}, [{\bf p}, {\bf S}]]^i\right]=-\frac{e}{mc}\left\{[{\bf B}, {\bf S}]-\frac{1}{2mc}[[{\bf p}, {\bf E},]{\bf S}]\right\}.
\end{eqnarray}
Using the definition (\ref{pfd.9}) on l.h.s., and replacing ${\bf S}={\bf S}_T+O(1/c^2)$ on the r.h.s., we obatain the Thomas equation.

Similar situation arises for a rotating body in general relativity. Here equations of motion for spin can be deduced either from the analysis of Einstein equations in multipole formalism \cite{Dixon1964}, or in geometric setting, assuming the Fermi-Walker  transport of spin-vector \cite{Synge_1964}. The two spins turn out to be different, and related by gravitational analogy of Eq. (\ref{pfd.9}), compare Eq. (193) in \cite{Kop_2019} with our (\ref{pfd.9.1}).

\section{Manifestly covariant version of the Pauli theory.}\label{mcv}

As we saw above, in $1/c^2$\, -approximation the expression (\ref{mch.7})
leads to the Pauli theory and hence may be taken as the Hamiltonian of its manifestly covariant version. The Hamiltonian should be accompanied with Poincare-covariant generalization of the brackets (\ref{pfd.5.1})-(\ref{pfd.5.4}). Besides, the brackets should lead to $U(1)$\,-invariant equations of motion, so they must be invariant under the gauge transformation (\ref{mch.1.1}). The canonical brackets $\{x^\mu, p_\nu\}=\delta^\mu{}_\nu$, $\{p_\mu, p_\nu\}=\{p_\mu, S^{\alpha\beta}\}=0$ are not invariant, and should be properly modified. For instance, if we substitute (\ref{mch.1.1}) into the bracket $\{x^\mu, p_\nu\}=\delta^\mu{}_\nu$, we obtain, $\{x^\mu, p'_\nu\}=\delta^\mu{}_\nu+\frac{e}{2c{\cal P}^2}S^{\mu\alpha}\partial_\alpha\partial_\nu\lambda$, instead of $\{x^\mu, p'_\nu\}=\delta^\mu{}_\nu$, where the extra-contribution is due to the  relativistic generalization (\ref{mcv.2}) of Eq. (\ref{pfd.5.1}).

We propose the following set of Poincare-covariant and $U(1)$\,-invariant brackets
\begin{eqnarray}
\{x^\mu, x^\nu\}=-\frac{1}{2{\cal P}^2}S^{\mu\nu}, \quad \qquad \quad \qquad\quad \qquad \quad \qquad \quad \label{mcv.2} \\
\{x^\mu, p_\nu\}=\delta^\mu{}_\nu-\frac{e}{2c{\cal P}^2}S^{\mu\alpha}\partial_\alpha A_\nu, \qquad \qquad \qquad\qquad \label{mcv.3}  \\
\{x^\alpha, S^{\mu\nu}\}=\frac{1}{{\cal P}^2}{\cal P}^{[\mu}S^{\nu]\alpha},  \qquad \qquad \qquad\qquad\qquad\qquad \label{mcv.4} \\
\{p_\alpha, S^{\mu\nu}\}=\frac{e}{c{\cal P}^2}{\cal P}^{[\mu}S^{\nu]\beta}\partial_\alpha A_\beta,  \qquad \qquad \qquad\qquad\quad\label{mcv.5} \\
\{p_\mu, p_\nu\}=-\frac{e^2}{2c^2{\cal P}^2}S^{\alpha\beta}\partial_\alpha A_\mu\partial_\beta A_\nu, \qquad \qquad \qquad\qquad \label{mcv.6} \\
\{S^{\alpha\beta}, S^{\mu\nu}\}=2N^{(\alpha\mu}S^{\beta\nu)}+\frac{e}{c({\cal P}^2)^2}{\cal P}^{[\alpha}(SFS)^{\beta][\mu}{\cal P}^{\nu]}. \label{mcv.7}
\end{eqnarray}
For the canonical momenta ${\cal P}_\mu=p_\mu-\frac{e}{c}A_\mu$ they imply
\begin{eqnarray}\label{mcv.8}
\{x^\mu, {\cal P}_\nu\}=\delta^\mu{}_\nu, \qquad \{{\cal P}^\mu, {\cal P}^\nu\}=\frac{e}{c}F^{\mu\nu}, \qquad \{S^{\mu\nu}, {\cal P}_\alpha\}=\frac{e}{c{\cal P}^2}{\cal P}^{[\mu}S^{\nu]\beta}F_{\beta\alpha}.
\end{eqnarray}
Computing $\dot z^A=\{z^A, H\}$, we obtain the Hamiltonian equations
\begin{eqnarray}
\dot x^\mu=\frac{1}{m}{\cal P}^\mu-\frac{e\mu}{2mc{\cal P}^2}(SF{\cal P})^\mu+\frac{e\mu}{8mc{\cal P}^2}S^{\mu\alpha}\partial_\alpha(FS), \qquad \qquad \qquad \qquad \qquad \qquad \qquad \label{mcv.9} \\
\dot{\cal P}^\mu=\frac{e}{mc}F^{\mu\nu}{\cal P}_\nu-\frac{e^2\mu}{2mc^2{\cal P}^2}(FSF{\cal P})^\mu+\frac{e\mu}{4mc}\partial^\mu(FS), \qquad \qquad \qquad \qquad \qquad \qquad \qquad \label{mcv.10} \\
\dot S^{\mu\nu}=\frac{e\mu}{mc}F^\mu{}_\alpha S^{\alpha\nu}+\frac{e(1-\mu)}{mc{\cal P}^2}{\cal P}^\mu(SF{\cal P})^\nu
+\frac{e\mu}{4mc{\cal P}^2}{\cal P}^\mu S^{\nu\alpha}\partial_\alpha(FS)-\frac{e^2\mu}{2mc^2({\cal P}^2)^2}{\cal P}^\mu(SFSF{\cal P})^\nu-(\mu\leftrightarrow\nu) \nonumber \\
\equiv\frac{e\mu}{mc}F^\mu{}_\alpha S^{\alpha\nu}+2{\cal P}^\mu\left[\dot x^\nu+\frac{e}{2mc{\cal P}^2}(SF{\cal P})^\nu-\frac{e^2\mu}{4mc^2({\cal P}^2)^2}(SFSF{\cal P})^\nu\right]-(\mu\leftrightarrow\nu). \qquad \qquad \qquad \qquad \label{mcv.11}
\end{eqnarray}
Together with the algebraic equations $S^{\mu\nu}{\cal P}_\nu=0$, $H=0$, they give manifestly Poincare-covariant version of the Pauli theory for an arbitrary value of magnetic moment $\mu$. The brackets (\ref{mcv.2})-(\ref{mcv.7}) obey the condition (\ref{mcf.4}), so the spin supplementary condition  $S^{\mu\nu}{\cal P}_\nu=0$ is consistent with the dynamical equations (\ref{mcv.9})-(\ref{mcv.11}). The mass-shell condition $H=0$ is consistent by construction: $\dot H=\{H, H\}=0$.

The Hamiltonian (\ref{mch.7}), being a linear function of spin and electromagnetic field strength, nevertheless gives the nonlinear equations of motion. This is due to the fact that an essential part of interaction turns out to be encoded in noncommutative brackets (\ref{mcv.2})-(\ref{mcv.8}).

In $1/c^2$\,-approximation our equations imply the equations (\ref{pau.7}) and (\ref{mch.4}).
As we saw above, they also represent equations of motion of the Hamiltonian theory (\ref{mch.8}). This proves that (\ref{mch.8}) is the physical-time Hamiltonian of the covariant theory (\ref{mch.7}) in $1/c^2$\,-approximation.

The equation (\ref{mcv.9}) in $1/c^3$\,-approximation implies the following expression for canonical momentum: ${\cal P}^\mu=m\dot x^\mu-\frac{e\mu}{2mc^3}(SF\dot x)^\mu+\frac{e\mu}{8m^2c^3}S^{\mu\alpha}\partial_\alpha(FS)$. This can be used to  exclude ${\cal P}^\mu$ from Eqs. (\ref{mcv.10}) and (\ref{mcv.11}), again in $1/c^3$-approximation.  Keeping only the linear on $F^{\mu\nu}$\,-terms (the approximation studied by Frenkel \cite{Frenkel}), the resulting equations coincide with those of Frenkel
\begin{eqnarray}\label{mcv.12}
\frac{d}{d\tau}\left[(m-\frac{e}{4mc^3}(SF))\dot
x^\mu+\frac{e}{8m^2c^3}S^{\mu\alpha}\partial_\alpha(FS)\right]=\frac{e}{c}(F\dot x)^\mu+\frac{e}{4mc}\partial^\mu(FS),
\quad
\end{eqnarray}
\begin{eqnarray}\label{mcv.13}
\dot S^{\mu\nu} =\frac{e}{mc}\left[F^{[\mu}{}_\alpha S^{\alpha\nu]}-\frac{1}{4mc^2}\dot
x^{[\mu}S^{\nu]\alpha}\partial_\alpha(FS)\right]\,, \qquad S^{\mu\nu}\dot x_\nu=0.
\end{eqnarray}
By the way, we showed that the approximately covariant Frenkel equations can be made covariant by adding the terms that are no more than quadratic in spin and field strength.

\section{The role of Thomas spin-vector in the covariant scheme.}\label{thp}

In this section we return back to the analysis of covariant theory in $1/c^2$\,-approximation. We assume  that the vector $S^i=\frac14\epsilon^{ijk}S^{jk}$ describe the spin-axis in the laboratory system.  Eqs.  (\ref{mcv.9})-(\ref{mcv.11}) at  $1/c^2$\,-order reduce to the Frenkel equations (\ref{pau.7}) and  (\ref{mch.4}). As we saw in Sect. \ref{pfd}, they give the Pauli quantum mechanics without any appeal to the Thomas vector  or to the Thomas precession formula. To understand the role of Thomas vector in the covariant formalism, we examine the transformational properties of Frenkel spin in the passage from laboratory to comoving and instantaneous frames.

Comoving observer is a non inertial system in which the particle is always in its center. For the four-dimensional vector and tensor fields given along the particle trajectory, the transformation from laboratory to comoving frame  is determined \cite{Schiff1960.1, Jackson} by the matrix
\begin{eqnarray}\label{thp.1.0}
\Lambda^\mu{}_\nu({\bf v(\tau)})=\left(\Lambda^0{}_0=\gamma, \Lambda^0{}_i=\Lambda^i{}_0=-\frac{\gamma}{c}v^i,
\Lambda^i{}_j=\delta^i{}_j+\frac{\gamma-1}{{\bf v}^2}v^iv^j\right),
\end{eqnarray}
where the functions $v^i(\tau)=\frac{dx^i(t)}{dt}|_{t(\tau)}$ represent velocity of the particle.
Using Eq. (\ref{mch.2}), it is convenient to write the velocity through the momenta, ${\bf v}(\tau)=c\frac{\boldsymbol{\cal P}}{{\cal P}^0}=\frac{{\bf p}(\tau)}{m}+O(1/c)$.  Using the transformation law of Frenkel
spin-tensor $S''^{\mu\nu}(\tau)=\Lambda^\mu{}_\alpha\Lambda^\nu{}_\beta S^{\alpha\beta}(\tau)$ and Eq. (\ref{mcf.3}),
we obtain for spatial components
\begin{eqnarray}\label{thp.1}
S'^{ij}=2\epsilon^{ijk}\left[S^k+\left(\frac{\gamma}{({\cal P}^0)^2}-\frac{\gamma-1}{(\boldsymbol{\cal P})^2}\right)[{\boldsymbol{\cal P}}, [{\boldsymbol{\cal P}}, {\bf S}]]^k\right].
\end{eqnarray}
We assume that the functions $S'^i_{comov}=\frac14\epsilon^{ijk}S'^{jk}$ describe the position of spin-axis in the comoving frame.
At the order $1/c^2$, Eq. (\ref{thp.1}) implies the following transformation law
\begin{eqnarray}\label{thp.4}
{\bf S}'_{comov}={\bf S}+\frac{1}{2(mc)^2}[{\bf p}, [{\bf p}, {\bf S}]].
\end{eqnarray}
Comparing this with the definition of Thomas spin (\ref{pfd.9}),  we see that accidentally, Thomas vector represents the spin axis  in comoving frame
\begin{eqnarray}\label{thp.4.1}
{\bf S}_T={\bf S}'_{comov}.
\end{eqnarray}
This clarifies the meaning of Thomas vector in the covariant theory: using the laboratory values ${\bf S}$ and ${\bf p}$, the laboratory observer can compute the vector ${\bf S}_T$ according to Eq. (\ref{pfd.9}), thus obtaining the magnitude and direction of spin-axis as it measured in comoving frame. From the equality  (\ref{thp.4.1}) we expect that the Thomas equation (\ref{pau.2}) describe the evolution of spin axis in the comoving frame. This can be confirmed by direct calculations. Derivative of Eq. (\ref{thp.4}) gives us the variation rate of spin-axis in comoving frame through the laboratory quantities as follows\footnote{Using the conversion factor $d/d\tau=\gamma d/dt=d/dt+O(1/c^2)$, we could obtain the variation rate with respect to the laboratory time. As we work in $1/c^2$\,-approximation, almost all equations of this section remain valid if we just replace $\frac{d}{d\tau}$ on $\frac{d}{dt}$.}:
\begin{eqnarray}\label{thp.5}
\frac{d{\bf S}'_{comov}}{d\tau}=\frac{d{\bf S}}{d\tau}-\frac{1}{2(mc)^2}[[{\bf p}, \frac{d{\bf p}}{dt}], {\bf
S}]+\frac{1}{(mc)^2}[{\bf p}, [\frac{d{\bf p}}{dt}, {\bf S}]].
\end{eqnarray}
Using Eqs. (\ref{pau.7}) and (\ref{mch.2}),  we obtain the equality
\begin{eqnarray}\label{thp.6}
\frac{d{\bf S}'_{comov}}{d\tau}=-\frac{e}{mc}\left\{[{\bf B}, {\bf S}]-\frac{1}{2mc}[[{\bf p}, {\bf E}],{\bf S}]\right\}.
\end{eqnarray}
If the dynamics of laboratory quantities ${\bf p}(\tau)$ and ${\bf S}(\tau)$ is known, Eq. (\ref{thp.6}) acquires the form $d{\bf S}'_{comov}/d\tau={\bf f}(\tau)$, with known function ${\bf f}(\tau)$.  Solving this equation, the laboratory observer will obtain the dynamics of spin axis as it seen by the comoving observer.
On the r.h.s we can replace ${\bf S}$ on ${\bf S}'_{comov}+O(1/c^2)$, thus obtaining the equation
\begin{eqnarray}\label{thp.6.1}
\frac{d{\bf S}'_{comov}}{d\tau}=-\frac{e}{mc}\left\{[{\bf B}, {\bf S}'_{comov}]-\frac{1}{2mc}[[{\bf p}, {\bf E}],{\bf S}'_{comov}]\right\}.
\end{eqnarray}
If the laboratory quantities ${\bf B}$, ${\bf E}$ and ${\bf p}(\tau)$ are  known, the laboratory observer can solve it for ${\bf S}'_{comov}(\tau)$.

The equation  (\ref{thp.6.1}) just coincides with the Thomas equation (\ref{pau.2}). This explains the meaning of Thomas equation in the covariant theory: solving Eq. (\ref{pau.2}) with use of laboratory quantities, the laboratory observer will obtain the functions $S^i_T(\tau)$, that describe the evolution of spin axis as seen by the comoving observer. We stress once again, that this interpretation is valid only in $1/c^2$\,-approximation.

For an electromagnetic field in comoving system we can write \cite{Jackson}
\begin{eqnarray}\label{thp.8}
{\bf E}={\bf E}'-\frac{1}{mc}[{\bf p}, {\bf B}']+O(1/c^2), \qquad {\bf B}={\bf B}'+\frac{1}{mc}[{\bf p}, {\bf
E}']+O(1/c^2).
\end{eqnarray}
Using Eqs. (\ref{thp.8}) and (\ref{mch.2}) in (\ref{thp.6}), we obtain the expression that can be thought as an equation of motion of spin axis in the comoving frame
\begin{eqnarray}\label{thp.11}
\frac{d{\bf S}'_{comov}}{d\tau}=-\frac{e}{mc}[{\bf B}', {\bf S}'_{comov}]+[{\boldsymbol\omega}_T, {\bf S}'_{comov}],
\end{eqnarray}
where
\begin{eqnarray}\label{thp.7.1}
{\boldsymbol\omega}_T=-\frac{1}{2(mc)^2}[{\bf p}, \frac{d{\bf p}}{dt}],
\end{eqnarray}
is the angular velocity vector of Thomas precession.  Comoving observer will detect torque of
spin exerted by magnetic field ${\bf B}'$, and an extra torque around the vector ${\boldsymbol\omega}_T$ due to non inertial character of
the comoving frame.

If ${\bf E}=0$ in the laboratory frame, Eq. (\ref{mch.2}) implies $d{\bf p}/dt=O(1/c)$, and (\ref{thp.11}) reduces to the equation for precession of spin in a magnetic field
\begin{eqnarray}\label{thp.10}
\frac{d{\bf S}'_{comov}}{d\tau}=-\frac{e}{mc}[{\bf B}', {\bf S}'_{comov}].
\end{eqnarray}

To complete the analysis, we dscuss the Frenkel vector in the instantaneous frame.  Instantaneous at the instant $\tau_0$ rest frame  (instantaneous frame for short), is the inertial system obtained by Lorentz boost (\ref{thp.1.0}),  where the numbers $v^i$ are equal to the velocity of a particle at fixed instant $\tau_0$.  Using the transformation law of Frenkel
spin-tensor $S''^{\mu\nu}(\tau)=\Lambda^\mu{}_\alpha\Lambda^\nu{}_\beta S^{\alpha\beta}(\tau)$ and Eq. (\ref{mcf.3}),
we obtain for spatial components
\begin{eqnarray}\label{thp.1.1}
S''^{ij}=2\epsilon^{ijk}\left[S^k-\frac{\gamma}{c{\cal P}_0}[{\bf v}, [{\boldsymbol{\cal P}}, {\bf S}]]^k-\frac{\gamma-1}{{\bf v}^2}[{\bf v},
[{\bf v}, {\bf S}]]^k\right].
\end{eqnarray}
At the order $1/c^2$, this implies the following transformation law of the Frenkel and Thomas spin-vectors
\begin{eqnarray}\label{thp.2.2}
{\bf S}''_{inst}={\bf S}+\frac{1}{mc^2}[{\bf v}, [{\bf p}, {\bf S}]]-\frac{1}{2c^2}[{\bf v}, [{\bf v}, {\bf S}]].
\end{eqnarray}
\begin{eqnarray}\label{thp.2.3}
{\bf S}''_{T, inst}={\bf S}_T+\frac{1}{2mc^2}[[{\bf v}, {\bf p}], {\bf S}_T].
\end{eqnarray}
Here ${\bf v}$ is velocity of the particle at $\tau_0$, while ${\bf p}$ is its momentum at the instant of observation. Both Thomas (\ref{pau.2}) and Frenkel (\ref{pau.7})  equations preserve their form under these transformations, and thus can be used as the equations of motion for these quantities by any inertial observer.
Computing derivative of these expressions, we obtain the variation rates in instantaneous frame through the laboratory quantities
\begin{eqnarray}\label{thp.3}
\frac{d{\bf S}''_{inst}}{d\tau}=\frac{d{\bf S}}{d\tau}+\frac{1}{mc^2}[{\bf v}, [\frac{d{\bf p}}{dt}, {\bf S}]].
\end{eqnarray}
\begin{eqnarray}\label{thp.3.1}
\frac{d{\bf S}''_{T, inst}}{d\tau}=\frac{d{\bf S}_T}{d\tau}+\frac{1}{2mc^2}[[{\bf v},  \frac{d{\bf p}}{dt}], {\bf S}_T].
\end{eqnarray}
At the instant $\tau_0$ the particle is instantanaously at rest, so ${\bf p}={\bf v}$, and Eq. (\ref{thp.2.2}) coincides with (\ref{thp.4}), as it is expected. Eq. (\ref{thp.2.3}) at this instant gives ${\bf S}''_{T, inst}(\tau_0)={\bf S}_T(\tau_0)$.

Taking the difference of equations (\ref{thp.5}) and (\ref{thp.3}) at the instant $\tau_0$,  and replacing in the resulting expression
${\bf S}={\bf S}''_{inst}+O(1/c^2)$, we relate the variation rates of spin axis in comoving and instantaneous frames, thus obtaining the
Thomas precession of Frenkel spin
\begin{eqnarray}\label{thp.7}
\frac{d{\bf S}'_{comov}}{d\tau}-\frac{d{\bf S}''_{inst}}{d\tau}=[{\boldsymbol\omega}_T, {\bf S}''_{inst}].
\end{eqnarray}
The difference at the instant $\tau_0$ is a rotation of spin axis ${\bf S}''_{inst}$ around the vector (\ref{thp.7.1}), with the angular velocity
equal to its length. So the non inertial comoving frame looks rotating in the system of instantaneous observer. Thomas
explained  this kinematic effect analysing the product of Lorentz boosts \cite{Thomas1927}.

We emphasize that formally similar equations (\ref{pfd.9}), (\ref{thp.4}) and (\ref{thp.2.2}) (the latter taken at $\tau_0$) represent $1/c^2$\,-approximation of different equations, and so they have a completely different meaning. The inaccuracies (including inaccuracies in notation) made by different authors in the derivation and analysis of the equations like (\ref{pfd.9}), (\ref{thp.4}) and (\ref{thp.2.2}), represent the source
of numerous confusion in the literature. Detailed analysis of the controversial works on the subject was undertaken in \cite{stepanov2012, Malykin_2007, Ritus_2007}. Here we clarified the meaning of these equations in the framework of covariant theory (\ref{mcv.9})-(\ref{mcv.11}).

\section{Conclusion.}\label{con}

In this work, we presented the manifestly covariant version of Pauli theory for the description of a spinning electron in Hamiltonian formalism. The covariant Hamiltonian (\ref{mch.7}) is a linear function of Frenkel spin and of field strength. Covariant brackets of the theory have been obtained from the requirement of consistency of Hamiltonian equations with the spin supplementary condition $S^{\mu\nu}{\cal P}_\nu=0$. This implies rather nontrivial  deformation (\ref{mcv.2})-(\ref{mcv.7}) of the canonical brackets (\ref{pau.4}). In particular, the bracket $\{x^\mu, x^\nu\}=-\frac{1}{2{\cal P}^2}S^{\mu\nu}$ states that position variables are noncommutative, and the spin-induced noncommutativity survives even in the non interacting theory of a spinning electron.  The same bracket appeared in the Ghosh model of anyon \cite{Ghosh_1994}. Non relativistic spinning particle has the commuting position variables, see Sect. 5C in \cite{AAD_Rec}. This also follows from the above mentioned bracket, since $1/{\cal P}^2\sim 1/(mc)^2\rightarrow 0$ as $c\rightarrow\infty$. In other words, the spin-induced noncommutativity is a relativistic effect.  Hence the manifestly covariant description of spin inevitably leads to the theory endowed with position-position noncommutative geometry. As we have shown, it is this bracket that is responsible for transforming the covariant Hamiltonian into the Hamiltonian of Pauli.
In this regard, we point out that spinning particles represent an exceptional example of intrinsically noncommutative and relativistic-invariant theory, with the spin-induced noncommutativity that manifests itself already at the Compton scale. The effects due to noncommutative geometry are of considerable interest in the current
literature \cite{sheikh2001, Harco_2019, Kai_2018_1, Daszkiewicz15, Daszkiewicz16, Kai_2019, Kai_2018_2, Stech_2019, Kai_2018_3}, and certainly deserve a detailed study in the relativistic-invariant  context of spin-induced noncommutativity.

The brackets (\ref{mcv.2})-(\ref{mcv.7})  encode an essential part of spin-field interaction, and lead to the equations of motion (\ref{mcv.9})-(\ref{mcv.11})  quadratic on spin and  field strength. Hamiltonian equations consistent with a set of algebraic constraints could be obtained in a more systematic way by constructing a proper variational problem. The search for a variational problem for spinning particle has an almost centenary history, see for example \cite{Bal_1983, Kow_2006, AAD_Rec} and references therein. One possibility is to consider the spin-tensor as a composite object constructed from  non-Grassmann vector and its conjugated momentum, $S^{\mu\nu}=2(\omega^\mu\pi^\nu-\omega^\nu\pi^\mu)$.  The Lagrangian that, besides the dynamical equations implies all the necessary constraints, has been recently  proposed in \cite{AAD_2014, Freidel_2016}. The spin supplementary condition $S^{\mu\nu}{\cal P}_\nu=0$ arises here as a consequence of two second-class constraints imposed on the basic vector and its momentum. The constraints can be taken into account with help of Dirac bracket. The point here is that this leads to higher complicated (nonpolynomial on $S$ and $F$) equations of motion \cite{AAD_2014, Freidel_2016}. So, the existence of quadratic on $S$ and $F$ equations (\ref{mcv.9})-(\ref{mcv.11}) which are  consistent with the spin supplementary condition, seem to be rather nontrivial fact that deserve further investigation.  

In $1/c^2$\,-approximation, both the Hamiltonian (\ref{mch.8}) and brackets (\ref{pfd.5.1})-(\ref{pfd.5.4}) of the covariant theory differ from that of Pauli theory (\ref{pau.3}), (\ref{pau.4}).  Nevertheless, they lead to the same quantum mechanics without any appeal to the Thomas precession formula (\ref{thp.7}).
The second equation from (\ref{pfd.1}) shows, that position of the particle in quantum mechanics is given by the Pryce (d) operator \cite{pryce1948mass}. At the classical level, Pauli theory can be thought as $1/c^2$\,-approximation of the covariant theory written in special variables  (\ref{pfd.8}) and  (\ref{pfd.9}).  This observation explains the discrepancy between the covariant (\ref{mch.8}) and Pauli (\ref{pau.3}) Hamiltonians.

Accidentally, the relation (\ref{pfd.9}) between the Thomas and Frenkel spin-vectors in laboratory frame coincides with the transformation law of Frenkel spin-vector in the passage from laboratory to comoving frame, see (\ref{thp.4}). This clarifies the meaning of Thomas spin and of the Thomas equation in the covariant scheme: solving Eq. (\ref{pau.2}) with the use of the laboratory quantities, the laboratory observer will obtain the functions $S^i_T(\tau)$, that describe the evolution of spin axis as it seen by the comoving observer.

It would be interesting to apply the developed formalism, considering the three-dimensional spin as a spatial part of the four-dimensional spin-vector $S^\mu$ instead of the Frenkel spin-tensor $S^{\mu\nu}$. This could give a Hamiltonian formulation for the Bargmann-Michel-Telegdi equations \cite{BMT_1959}.

{\bf Acknowledgments.} The work of A.A.D has been supported by the Brazilian foundation CNPq (Conselho Nacional de
Desenvolvimento Cient\'ifico e Tecnol\'ogico - Brasil),  and by Tomsk State University Competitiveness Improvement
Program.
The work of D.M.T has been supported by  Coordena\c{c}\~{a}o de
Aperfei\c{c}oamento de Pessoal de N\'ivel Superior - Brasil (CAPES) -
Finance Code 001.


\begin{thebibliography}{99}

\bibitem{Gousmith_1926}
G. E. Uhlenbeck and G. E. Goudsmit, {\it Spinning electrons and structure of spectra}, Nature {\bf 117} (1926) 264.

\bibitem{Thomas1927}
L. H. Thomas, \textit{The kinematics of an electron with an axis}, Philosophical Magazine and Journal of Science {\bf
3} S.7, No.13 (1927) 1.

\bibitem{Frenkel} J. Frenkel, \textit{Die elektrodynamik des rotierenden elektrons}, Z. Phys. {\bf 37} (1926) 243.

\bibitem{Frenkel2}
J. Frenkel, {\it Spinning Electrons}, Nature {\bf 117} (1926) 653.

\bibitem{Field:1979}
J. H. Field, E. Picasso and F. Combley, {\em Tests of fundamental physical theories from measurements of free charged
leptons}, Sov. Phys. Uspekhi, {\bf 22} (1979) 199 (in Russian).

\bibitem{miller2007muon}
J.~P.\ Miller, E.\ de~Rafael, and B.~L.\ Roberts.
\newblock {\it Muon (g-2): experiment and theory},
\newblock  Reports on Progress in Physics, {\bf 70(5)} (2007) 795; Preprint CPT-P07-2007.

\bibitem{Struck_2019}
J. Struckmeier, D. Vasak, A. Redelbach, P. Liebrich and  H. St\"{o}cker, {\it Pauli-type coupling between spinors and curved spacetime}, arXiv:1812.09669. 

\bibitem{Dayi_2015}
O.F. Dayi, E. Kilincarslan, {\it A semiclassical kinetic theory of Dirac particles and Thomas precession}, Phys. Lett. B 749 (2015) 119; arXiv:1508.00781. 

\bibitem{Khripl_2000}
A. A. Pomeransky, R. A. Senkov and I. B. Khriplovich, {\it Spinning relativistic particles in external fields},
Phys.\ Usp.\  {\bf 170} (2000)1055;   [Usp.\ Fiz.\ Nauk {\bf 43} (2000) 1129].

\bibitem{Oancea_2019}
M. A. Oancea, C. F. Paganini, J. Joudioux and L. Andersson, {\it An overview of the gravitational spin Hall effect}, arXiv:1904.09963.

\bibitem{Gerakopoulos_2019}
V. Witzany, J. Steinhoff, G. Lukes-Gerakopoulos, {\it Hamiltonians and canonical coordinates for spinning particles in curved space-time}, Class. Quantum Grav. {\bf 36} (2019) 075003; arXiv:1808.06582.

\bibitem{Bosso_2019}
Pasquale Bosso and Saurya Das, {\it Lorentz invariant mass and length scales}, Int. J. Mod. Phys. {\bf D 28} (2019) 1950068;  	arXiv:1812.05595.

\bibitem{Nica_2019}
U. Nucamendi, R. Becerril and P. Sheoran, {\it Bounds on spinning particles in their innermost stable circular orbits around rotating braneworld black hole},  arXiv:1910.00156. 


\bibitem{Dehghani_2018}
Salman Abarghouei Nejad, Mehdi Dehghani, Majid Monemzadeh, {\it Spinning Toroidal Brane Cosmology; A Classical and Quantum Survey},  	arXiv:1901.03292.



\bibitem{Nagar_2018_25}
Alessandro Nagar, Sebastiano Bernuzzi, Walter Del Pozzo, Gunnar
Riemenschneider, Sarp Akcay, Gregorio Carullo, Philipp Fleig, Stanislav
Babak, Ka Wa Tsang, Marta Colleoni, Francesco Messina, Geraint Pratten,
David Radice, Piero Rettegno, Michalis Agathos, Edward Fauchon-
Jones, Mark Hannam, Sascha Husa, Tim Dietrich, Pablo Cerda-Duran,
Jose A. Font, Francesco Pannarale, Patricia Schmidt, Thibault Damour,
{\it Time-domain effective-one-body gravitational waveforms for coalescing
compact binaries with nonprecessing spins, tides and self-spin effects},  Phys. Rev. {\bf D 98} (2018) 104052;
arXiv:1806.01772.

\bibitem{Kaye_2019}
Kaye Jiale Li, Kinwah Wu, Dinesh Singh, {\it Spin dynamics of a millisecond pulsar orbiting closely around a massive black hole}, Mon. Not. R. Astron. Soc. {\bf 485}  (2019) 1053; 
arXiv:1902.03146.

\bibitem{Bosso_2018}
Vasil Todorinov, Pasquale Bosso and Saurya Das, {\it  Relativistic Generalized Uncertainty Principle}, Ann. Phys. (Amsterdam) {\bf 405} (2019) 92;  arXiv:1810.11761.

\bibitem{Peng_2019}    
Yu-Peng ZhangShao-Wen, WeiPau Amaro-Seoane, Jie Yang and Yu-Xiao Liu, {\it Motion deviation of test body induced by spin and cosmological constant in extreme mass ratio inspiral binary system}, Eur. Phys. Journal C {\bf 79} (2019) 856 (11). 

\bibitem{Pauli_1927}
W. Pauli, {\it On the quantum mechanics of magnetic electrons}, Zeit. f. Phys. {\bf 43} (1927) 601.

\bibitem{Dirac_1927}
P. A. M. Dirac, {\it The physical interpretation of the quantum dynamics}, Proc. Roy. Soc. (A) {\bf 113} (1927) 621.

\bibitem{foldy:1978}
L.~L. Foldy and S.~A. Wouthuysen, \textit{On the Dirac theory of spin 1/2 particles and its non-relativistic limit},
Phys. Rev. {\bf 78},  29 (1950)

\bibitem{Kramers_1934}
H. A. Kramers, {\it On the classical theory of the spinning electron}, Physica {\bf 1} (1934) 825.

\bibitem{Sak}
J. J. Sakurai, {\it Advanced quantum mechanics} (Addison-Wesley Publishing Company, 1967)

\bibitem{Jackson} J. D. Jackson, \textit{Classical Electrodynamics} (John Willey and Sons, 1975)

\bibitem{Weinb_QFT} S. Weinberg, \textit{The Quantum Theory of Fields}, vol. 1
(Cambridge University Press, Cambridge, 1995)

\bibitem{BMT_1959}
V. Bargmann, L. Michel and V.L. Telegdi, {\it Precession of the polarized particles moving in a homogeneous electromagnetic field}, Phys. Rev. Lett. {\bf 2} (1959) 435.

\bibitem{corben:1968}
H. C.\ Corben, {\it Classical and Quantum Theories of Spinning Particles}, (Holden-Day, San Francisco, 1968)

\bibitem{stepanov2012}
S. S.\ Stepanov, {\it Thomas Precession for spin and for a rod}, Physics of Particles and Nuclei {\bf 43} (2012) 128.

\bibitem{Malykin_2007}
G. B. Malykin, {\it Thomas precession: correct and incorrect solutions}, Uspekhi Fizicheskikh Nauk, {\bf 176 (8)} (2006) 865.

\bibitem{Ritus_2007}
V. I. Ritus, {\it On the difference between Wigner's and Moller's approaches to description of the Thomas precession}, Uspekhi Fizicheskikh Nauk, {\bf 177 (1)} (2007) 105. 

\bibitem{deriglazov2010classical}
A. Deriglazov, {\em Classical Mechanics: Hamiltonian and Lagrangian Formalism} (Springer, 2nd edition, 2017).

\bibitem{Git_2017}
J. Assirati, D. M. Gitman, {\it Covariant quantizations in plane and curved spaces}, Eur. Phys. J. C 77 (2017) 476; arXiv:1705.09960.

\bibitem{Gueor_2019}
V. G. Gueorguiev, {\it Reparametrization-Invariance and Some of the Key Properties of Physical Systems}, arXiv:1903.02483.

\bibitem{Hori_2019}
Takayuki Hori, {\it Another counter-example to Dirac's conjecture}, arXiv:1902.09296.

\bibitem{Ottinger_2018}
C. \"{O}ttinger, {\it Hamiltonian formulation of a class of constrained fourth-order differential equations in the Ostrogradsky framework}, J. Phys. Commun. {\bf 2} (2018) 125006, 1-10; arXiv:1810.02193.

\bibitem{Pavsic_2019}
Matej Pav\v{s}i\v{c}, {\it On the direct quantization of gravity coupled to matter and the emergence of time}, arXiv:1906.03987.

\bibitem{Breban_2018}
R. Breban, {\it The Four Dimensional Dirac Equation in Five Dimensions}, Annalen der
Physik, {\bf 530} (2018) p. 1800042; arXiv:1801.06054.

\bibitem{Darvas_2018}
G. Darvas, {\it Algebra of hypersymmetry (extended version) applied to state transformations in strongly relativistic interactions illustrated on an extended form of the Dirac equation}, arXiv:1809.05396.

\bibitem{Ghosh_1994} 
S. Ghosh, {\it Spinning particles in $2+1$ dimensions}, Phys. Lett. B 338, 235, (1994); arXiv:hep-th/9406089. 

\bibitem{DPM2016}
A. A. Deriglazov and A. M. Pupasov-Maksimov, {\it Relativistic corrections to the algebra of position variables and
spin-orbital interaction}, Phys. Lett. {\bf B 761} (2016)  207.

\bibitem{AAD_Rec}
Alexei A. Deriglazov and Walberto Guzm\'an Ram\'irez, {\it Recent progress on the description of relativistic spin: vector model of spinning particle and rotating body with gravimagnetic moment in General Relativity}, Advances in Mathematical Physics, Volume 2017 (2017), Article ID 7397159, 49 pages; arXiv:1710.07135.

\bibitem{Dixon1964}
W. G. Dixon, \textit{A Covariant Multipole Formalism for Extended Test Bodies in General Relativity},  Nuovo Cimento
{\bf 34} (1964) 317.

\bibitem{Synge_1964}
J. L. Synge. {\it Relativity: the general theory}, (Series in Physics,  North-Holland, Amsterdam, 1964)

\bibitem{Kop_2019}
Sergei M. Kopeikin, {\it Covariant Equations of Motion of Extended Bodies with Arbitrary Mass and Spin Multipoles},  Phys. Rev.  {\bf D 99}  (2019) 084008; arXiv:1810.11713.

\bibitem{Schiff1960.1}
L. I. Schiff, {\it Motion of a gyroscope according to Einstein's theory of gravitation},
Proc. Natl. Acad. Sci. U. S. {\bf 46} (1960)  871. 

\bibitem{sheikh2001}
M. Chaichian, M. M. Sheikh-Jabbari and A. Tureanu, {\it Hydrogen atom spectrum and the lamb shift in noncommutative QED}, Phys. Rev. Lett. {\bf 86} (2001) 2716.

\bibitem{Harco_2019}
Tiberiu Harko, Shi-Dong Liang, {\it Energy-dependent noncommutative quantum mechanics}, Eur. Phys. J. {\bf C 79} (2019) 300;  arXiv:1903.06776.

\bibitem{Kai_2018_1}
Kai Ma, Ya-Jie Ren, Ya-Hui Wang, {\it Probing Noncommutativities of Phase Space by Using Persistent Charged Current and Its Asymmetry}, Phys. Rev. {\bf D 97} (2018) 115011 ; arXiv:1703.10923.

\bibitem{Daszkiewicz15}
M. Daszkiewicz, \textit{Generalized twist deformations of Poincar\'{e} and Galilei quantum groups}, Mod. Phys. Lett. {\bf A 30},  1550034 (2015).

\bibitem{Daszkiewicz16}
 M. Daszkiewicz, \textit{Photoelectric effect for twist-deformed space-time}, Acta Phys.Polon. {\bf B 47}, 1293 (2016) 

\bibitem{Kai_2019}
Kai Ma, {\it Chiral Spin Noncommutative Space and Anomalous Dipole Moments}, arXiv:1903.11439.

\bibitem{Kai_2018_2}
Ya-Jie Ren, Kai Ma, {\it Influences of the coordinate dependent noncommutative space on charged and spin currents}, Int. J. Mod. Phys. {\bf A 33} (2018) 1850093; arXiv:1802.10452.

\bibitem{Stech_2019}
C. A. Stechhahn, {\it Aharonov-Bohm scattering for relativistic particles in (3 + 1)-dimensional noncommutative space with spin dependence},  arXiv:1905.12538.

\bibitem{Kai_2018_3}
Kai Ma, Jian-Hua Wang, Huan-Xiong Yang,  {\it Time-dependent Aharonov-Bohm effect on the noncommutative space}, Phys. Lett. {\bf B 759} (2016) 306; arXiv:1604.02110.

\bibitem{Bal_1983}   
A. P. Balachandran, G. Marmo, B. S. Skagerstam and A. Stern, {\it Gauge Theories and Fibre Bundles - Applications to Particle Dynamics}, Lecture Notes in Physics, 188 (1983); arXiv:1702.08910. 

\bibitem{Kow_2006}
L. Freidel, J. Kowalski-Glikman, A. Starodubtsev, {\it Particles as Wilson lines of gravitational field}, Phys. Rev. {\bf D 74} (2006) 084002; arXiv: gr-qc/0607014.   

\bibitem{AAD_2014}   
A. A. Deriglazov, {\it Lagrangian for the Frenkel electron}, Phys. Lett. {\bf B 736} (2014) 278; arXiv:1406.6715. 

\bibitem{Freidel_2016}
T. Rempel and  L. Freidel, {\it Interaction Vertex for Classical Spinning Particles}, Phys. Rev. {\bf D 94} (2016) 044011;  arXiv:1507.05826.

\bibitem{pryce1948mass}
M.~H.~L. Pryce, \textit{The mass-centre in the restricted theory of relativity and its connexion with the quantum
theory of elementary particles}, Proceedings of the Royal Society of London. Series A. Mathematical and Physical
Sciences {\bf 195} (1948) 62.

\end{thebibliography}
\end{document}